# Structural and Chemical Orders in $Ni_{64.5}Zr_{35.5}$ Metallic Glass by Molecular Dynamics Simulation

L. Tang[1,2], T. Q. Wen[2,3], N. Wang[3,*], Y. Sun[2], F. Zhang[2], Z. J. Yang[1,2], K. M. Ho[2,4], and C. Z. Wang[2,4,†]

[1]*Department of Applied Physics, College of Science, Zhejiang University of Technology, Hangzhou, 310023, China*

[2]*Ames Laboratory-USDOE, Iowa State University, Ames, Iowa 50011, USA*

[3]*MOE Key Laboratory of Materials Physics and Chemistry under Extraordinary Conditions, School of Natural and Applied Sciences, Northwestern Polytechnical University, Xi'an 710072, China*

[4]*Department of Physics and Astronomy, Iowa State University, Ames, Iowa 50011, USA*

Corresponding authors:   \* nan.wang@nwpu.edu.cn       † wangcz@ameslab.gov

**Abstract**

The atomic structure of $Ni_{64.5}Zr_{35.5}$ metallic glass has been investigated by molecular dynamics (MD) simulations. The calculated structure factors from the MD glassy sample at room temperature agree well with the X-ray diffraction (XRD) and neutron diffraction (ND) experimental data. Using the pairwise cluster alignment and clique analysis methods, we show that there are three types dominant short-range order (SRO) motifs around Ni atoms in the glass sample of $Ni_{64.5}Zr_{35.5}$, i.e., Mixed-Icosahedron(ICO)-Cube, Twined-Cube and icosahedron-like clusters. Furthermore, chemical order and medium-range order (MRO) analysis show that the Mixed-ICO-Cube and Twined-Cube clusters exhibit the characteristics of the crystalline B2 phase. Our simulation results suggest that the weak glass-forming ability (GFA) of $Ni_{64.5}Zr_{35.5}$ can be attributed to the competition between the glass forming ICO SRO and the crystalline Mixed-ICO-Cube and Twined-Cube motifs.

I. **Introduction**

Atomistic structures of metallic glasses (MG) have been attracted considerable attentions[1-8] since the extraordinary properties such as strength and elasticity in MG systems are closely related to the atomic packing structures in the glasses[9]. Although MGs do not exhibit long-range translational and rotational orders, experiments [10-12] and theoretical simulations [13, 14] have indicated that there are short- to medium-range orders in MGs. It is believed that the icosahedral (ICO) short-range order (SRO) is responsible for glass-forming ability (GFA) [15-17]. Meanwhile, SRO and medium-range order (MRO) also govern the mechanical response to deformation in the MG systems [18]. Hence, investigating such atomic SRO and MRO structures in metallic liquids and glasses is a key to understanding structure-properties relationship of MG materials.

Among all Ni-Zr alloys with different compositions, it is believed that $Ni_{64.5}Zr_{35.5}$ could be the most promising candidate for forming bulk MG, since $Ni_{64.5}Zr_{35.5}$ is close to a eutectic point in Ni-Zr binary phase diagram [19-21]. Because the melting temperature is at a local minimum at the eutectic composition, the kinetics associated with crystal nucleation and growth would be relatively easily suppressed, leading to a relatively higher GFA compared to the same alloy system at other compositions. In particular, it has been shown that critical cooling rate for glass formation is minimal for the Ni-Zr alloy containing about 35 at.% Zr [22], which suggests that $Ni_{64.5}Zr_{35.5}$ has the highest GFA in Ni-Zr system. However, compared to well-known Cu-Zr system which has been commonly regarded as good glass former in binary metallic alloys [23, 24], experimental synthesis of bulk MG of $Ni_{64.5}Zr_{35.5}$ is still very challenging.

Although Ni and Cu differ only by one electron, the GFA of $Cu_{64.5}Zr_{35.5}$ and $Ni_{64.5}Zr_{35.5}$ is very different. $Cu_{64.5}Zr_{35.5}$ glassy samples with mm size can be fabricated by copper mold casting method [23], while glassy $Ni_{64.5}Zr_{35.5}$ ribbon can only be obtained by careful melt-spinning process [20]. The cooling rate in melt-spinning method is typically $10^5$-$10^6$ K/s which is much faster (about 2 orders of magnitude) than that of casting method. This difference indicates that the GFA of $Ni_{64.5}Zr_{35.5}$ is much weaker relative to that of $Cu_{64.5}Zr_{35.5}$ system. This weaker GFA nature of $Ni_{64.5}Zr_{35.5}$ could be attributed to the difference in the short- to medium-range order in the atomistic structures between the two systems. For example, it is believed that crystal-like motif can facilitate the crystal nucleation and growth to prevent the glass formation if the cooling rate is not high enough. Therefore, investigating the differences of short- to medium-range ordered structure and chemical order between $Ni_{64.5}Zr_{35.5}$ and Cu-Zr system would provide useful insights into the origin of weak GFA in $Ni_{64.5}Zr_{35.5}$.

Using the experimental X-ray diffraction (XRD) and neutron diffraction (ND) experimental data, the atomic structure of $Ni_{64.5}Zr_{35.5}$ MG has been studied by reverse Monte-Carlo (RMC) method and Voronoi tessellation analysis [25, 26]. These studies have suggested that a large number of icosahedral-like and prismatic-like SRO structures exist in $Ni_{64.5}Zr_{35.5}$ glass. Moreover, a significant degree of chemical ordering has also been observed in $Ni_{64.5}Zr_{35.5}$ glass samples [26]. However, compared to the RMC method, MD simulation has a great advantage in obtaining more reliable atomic structure as long as the interatomic potentials are accurate. Fortunately, the interatomic potential of Ni-Zr system has been developed recently [27], which allows us to perform direct MD simulations for $Ni_{64.5}Zr_{35.5}$. To our knowledge, although the MD simulations for liquid Ni-Zr binary alloys have been reported [27], MD simulation of metallic glass at the composition around $Ni_{64.5}Zr_{35.5}$ is still lacking. Moreover, no detailed explanation of experimental results by MD simulation has been reported yet. Reliable MD simulation for this system will enable us to look into the structures of the alloy at undercooled liquid and glass states at atomistic level, and provide very useful insights into the GFA in this system.

In this paper, the atomic structure of $Ni_{64.5}Zr_{35.5}$ glass is studied by MD simulation using the Finnis-Sinclair-type potential recently developed [27, 28]. The pairwise cluster alignment method and clique analysis algorithm [29, 30] are used to identify the dominant structure order in MD $Ni_{64.5}Zr_{35.5}$ sample. This approach can extract a clique of similar clusters from atomic structure of MD sample without knowing the details of packing motif in advance. Meanwhile, the cluster alignment method can also calculate the populations of SRO clusters and explore the MRO in metallic liquid and glass systems. In addition to a large fraction of the Twined-Cube (similar to prismatic-like structure) and icosahedron-like clusters around Ni atoms as previously observed, we found significant populations of Ni-centered Mixed-ICO-Cube cluster which is composed of half icosahedron and half cube in $Ni_{64.5}Zr_{35.5}$ glass sample obtained from the MD simulations. Furthermore, we show that the cubic part of Mixed-ICO-Cube cluster and the Twined-Cube cluster have excellent chemical ordering, indicating the emergence of metastable crystalline B2 phase [31]. Our results show that the suppression of glass formation by crystalline B2 phase could be responsible for the weak GFA of $Ni_{64.5}Zr_{35.5}$.

## II. Method

The sample used in MD simulation of $Ni_{64.5}Zr_{35.5}$ contains 3225 Ni and 1775 Zr atoms. The simulations are performed using the isothermal-isobaric (NPT) ensemble with the Nosé-Hoover thermostat in LAMMPS code [28]. The semi-empirical Finnis-Sinclair-type interatomic potential developed by Wilson and Mendelev is employed [27]. The time step in the MD simulations is 2.5 fs. Before the cooling process, the sample is held at 2000K for 2.5ns to achieve equilibrium. After that, the sample is cooled down to 300K continuously with different cooling rates at $10^{13}$, $10^{12}$, $10^{11}$ and $10^{10}$ K/s, respectively. In order to eliminate the effect of atomic thermal motions, the structural and chemical orders in the glass sample at 300 K are averaged over 500ps, which is sufficient to obtain the convergent results for the structural and physical properties studied in this paper.

In order to compare with the experimental data, the total structure factors S(*q*) of the Ni$_{64.5}$Zr$_{35.5}$ MD samples are calculated by the Faber-Ziman formalism[32]

$$S(q) = \sum_{i \leq j} w_{ij} S_{ij}(q) \quad (1)$$

where $i$ or $j$ denotes atomic specie. Here the partial structure factor $S_{ij}(q)$ is

$$S_{ij}(q) = 1 + 4\pi\rho_j \int_0^\infty [g_{ij}(r) - 1] \frac{\sin(qr)}{r} r dr \quad (2)$$

where $g_{ij}(r)$ is the partial pair correlation function and $\rho_j$ is the number density of the relevant atom specie. To compare with the XRD results, the *q*-dependent scattering factors [33]

$$w_{ij}^{\text{XRD}}(q) = (2 - \delta_{ij}) c_i c_j \frac{f_i(q) f_j(q)}{[\sum_i c_i f_i(q)]^2} \quad (3)$$

are used in the Faber-Ziman formalism, where $c_i$ and $c_j$ are the compositions of the relevant atom species and *f$_i$(q)* and *f$_j$(q)* are the corresponding *q*-dependent atomic scattering factors in XRD experiment. For comparison with neutron diffraction (ND) data we use the weighting coefficients[26]

$$w_{ij}^{ND} = (2 - \delta_{ij})c_i c_j \frac{b_i b_j}{(\sum_i c_i b_i)^2} \qquad (4)$$

where $b_i$ and $b_j$ are the coherent neutron scattering length of the relevant atom species [34].

The SROs in $Ni_{64.5}Zr_{35.5}$ MD samples are analyzed by the pairwise cluster alignment method [29, 30]. In this method, a cluster can be assigned to each atom in the MD sample by extracting this atom and its first neighbor cell atoms along with it from the MD sample. Then each cluster extracted from the MD sample is used as a template and all other clusters with the same chemical element at the center are aligned against it. The similarity between the aligned cluster and the template is measured by an alignment score defined as

$$f = \min_{0.8 \leq \alpha \leq 1.2} \left[ \frac{1}{N} \sum_{i=1}^{N} \frac{(\vec{r}_{ic} - \alpha \vec{r}_{it})^2}{(\alpha \vec{r}_{it})^2} \right]^{1/2} \qquad (5)$$

where $N$ is the number of the neighbor atoms in the template. $\vec{r}_{ic}$ and $\vec{r}_{it}$ are the atomic positions in the aligned cluster and template, respectively. To obtain optimal alignment, $\alpha$ is chosen between 0.8 and 1.2 to vary the size of the template. Once we have all the alignment scores between any two clusters in the glass sample, we can obtain a clique of clusters whose pairwise alignment scores are smaller than a cutoff value of 0.15. The clusters belong to the same clique are very similar with each other and have the common SRO.

## III. Results and Discussions

Fig. 1 shows the evolution of instantaneous potential energy $E-3k_BT$ as function of temperature with a continuous cooling rate of $10^{10}$ K/s. In harmonic systems, the temperature dependence of both kinetic energy and potential energy are $3/2\ k_BT$. Therefore, $E-3k_BT$ will be a constant for all temperatures in harmonic systems. By subtracting $3k_BT$ from the total energy obtained from MD simulations, the temperature dependence of anharmonicity in the system can be seen more clearly [35]. In particular, the plot in Fig. 1 shows that the anharmonicity in liquid and glass are different, which enables us to see the glass transition and estimate the glass transition temperature $T_g$ from the plot. From Fig. 1 one can see that at temperature higher than 975K the energy decreases significantly with decreasing temperature, whereas at low temperature region with $T$<975K the energy has only a slight decrease and is nearly constant around room temperature. Clearly, it shows a transition from liquid to glass and the obtained glass transition temperature is $T_g \approx$975K, which agrees with the experimental result [19].

In order to see how good are the glass samples obtained from our MD simulations in comparison with experiment, we calculated the total and partial structure factors using Faber-Ziman formalism[32] for the samples at 300K prepared by different cooling rates. As shown in Figs. 2(a) and 2(b), although there is a slight shift of the height of the main peak, one can observe that the structure factors calculated from our MD simulations using the Eq. (3) and Eq. (4) agree well with the experimental XRD and ND data [26], respectively. It should be noted that as the same problem in RMC method, the agreement between the calculated and experimental S($q$) does not always guarantee the correct description of the three-dimensional atomistic structures in the glass sample. Nevertheless, this good agreement of total structure factors is an essential requirement of the reliable MD simulation.

We can see from Figs. 2(a) and (b) that the height of main peak increases with lowering cooling rate, indicating more order developed with lower cooling rate. In Figs. 2(b) the experimental ND data and the simulated ND spectrum exhibit a pre-peak [25] around $q \approx 1.8 \text{Å}^{-1}$ in addition to other peaks. Because the low $q$ region in reciprocal space is corresponding to large distance in real space, the pre-peak would indicate certain medium-range correlations in the glass structure. However, the pre-peak is not well seen in the in XRD spectra from both experiment and simulation. Therefore, the origin of this small pre-peak needs further investigation.

The comparisons between the calculated and experimental partial structure factors [26] are shown in Fig. 3. The overall agreements between the calculated and experimental spectra are reasonably good but some discrepancies present, especially for the partial S($q$) of the Zr-Zr pairs. Although the peak positions agree with each other, the peak intensity of the first peak from the calculation is much higher than that seen in experiment. It should be noted that the experimental partial structure factors are calculated from the different total structure factors of isotopic substitutions. The differences of total structure factors among the isotopic substitutions may be very small, so that the uncertainty of the obtained partial structure factors could be relative larger. Another source of the differences may also be attributed to the inaccuracy the interatomic potential used in the simulation, but we also would like to note that the potential has been successfully applied to explain several phenomena in $Ni_{50}Zr_{50}$ (see Ref. [36]) Nevertheless, the main discrepancy is in the intensity of the peak at small $q$ value which affects the degree of MRO in the system. The peaks of Ni-Ni and Ni-Zr around $q=3$ (Å$^{-1}$) corresponding to the dominant SRO clusters in Ni$_{64.5}$Zr$_{35.5}$ agree well with experiment, suggesting the SRO results from our simulation and analysis should be reliable.

The structure analysis based on the atomistic model from our MD simulations shows that the dominant motifs around Ni and Zr atoms in Ni$_{64.5}$Zr$_{35.5}$ glass at 300K are Mixed-ICO-Cube, Twined-Cube and ICO-like clusters. The corresponding structures of Mixed-ICO-Cube and Twined-Cube motifs are shown in Fig. 4. In Mixed-ICO-Cube motif, the center Ni atom is surrounded by 10 atoms. This motif can be viewed as a combination of a half cube and a half icosahedron as shown in Fig. 4(a). The sites (2, 3, 4, 7, 8, 9, 10) are the seven corners of a slightly distorted cube, forming three nearly orthogonal squares. On the other side, the center Ni atom is also enclosed by 10 triangles which form the surface of a half slightly distorted icosahedron. Similarly, the Twined-Cube motif can be regarded as two interpenetrating cubes C1 and C1' sharing a common (111) plane as shown in Fig. 4(b). Thus, the center Ni atom in Twined-Cube cluster is surrounded by 6 faces (11 atoms) which form two groups of nearly orthogonal squares. By contrast, ICO-like clusters in our MD samples (not shown) are nearly identical to the ideal ICO motif, with only slightly distortions.

After identifying the dominant structure motifs in the glass sample, we take these three motifs along with some common crystalline motifs such as BCC, FCC and HCP motifs as templates and perform cluster-template alignment for all the Ni-centered clusters extracted from the MD samples to calculate the alignment scores distribution against these templates. Since no dominant motif of Zr-centered clusters is found by the pairwise alignment method other than Z15 and Z16, and the coordination number of Zr in our sample is about 15.6, we only include the common motifs Z15 and Z16 clusters as templates to analyze the SRO of Zr-centered clusters. Figs. 5(a) and 5 (b) show the distribution of alignment scores for Ni- and Zr-centered clusters from MD glass sample at 300K, respectively. As shown in Fig. 5(a), compared to the other motifs of Ni-centered cluster, the distribution of scores for Mixed-ICO-Cube, Twined-Cube and ICO-like motifs have relatively larger portion in the region with small score. It indicates that the local structures in the sample can be better described by these motifs than the others. Similarly, Fig. 5(b) shows the Z15 cluster is the major motif of Zr-centered cluster in $Ni_{64.5}Zr_{35.5}$.

To describe the SRO of glass sample quantitatively, we use alignment score 0.15 as a cutoff to assign the clusters to the given template. If a cluster has alignment score less than 0.15 for more than one template, the lowest alignment score is used to assign the motif of the cluster. The choice of cutoff value of 0.15 for determining the fraction of the SRO motif is a bit arbitrary but it is guided by considering the thermal motion of the atoms at given temperature. This cutoff value is inferred from the width of the first peak in the calculated pair correlation g(r) which reflects the atomic displacements due to thermal motion. At 300K, the width of the first peak of g(r) is roughly 0.15 $d_0$, where $d_0$ is the averaged nearest neighbor bond length.

Table. I show the fractions of SROs in glass sample at 300K prepared by continuous cooling of $10^{10}$ K/s. As a result, we found that the fractions of Ni-centered Mixed-ICO-Cube, Twined-Cube, and ICO-like motifs in the $Ni_{64.5}Zr_{35.5}$ glass sample at 300K are 18.6%, 9.3% and 11.8% of the total Ni atoms, respectively. By contrast, the fraction of Zr-centered Z15 motif is only about 4.8% of total Zr atoms. Other kinds of Ni (Zr) centered motifs such as BCC, FCC and HCP have less than 5% of the total Ni (Zr) atoms. Since the populations of these minority motifs are much less than those of the dominant motifs around the Ni central atoms, such minority motifs will be ignored in the following discussions. From the alignment score distributions shown in Fig. 5, we can see that different choice of the cutoff value will change the fraction numbers of the motifs, but the relative ratios among different motifs are not sensitive to the choice of cutoff value as long as this value is reasonably chosen. Therefore, our conclusion that the Mixed-ICO-Cube, Twined-Cube, and ICO-like motifs are dominating the atomistic SRO in the glassy sample does not dependent on the choice of cutoff score.

Moreover, although only about 40% of the Ni atoms can be classified as the central atoms of the three dominant motifs under the cutoff score of 0.15, the total number of atoms involved in the first shells of these dominant clusters is about 95% of the total atoms in the sample. Therefore, the local structures in the system are dominated by the Ni-centered Mixed-ICO-Cube, Twined-Cube and ICO-like motifs. In fact, if we choose the cutoff value of 0.18 for the Ni-centered clusters, about 86% of Ni atoms can be assigned and the relative ratios among the above three motifs are almost the same as that with the cutoff score of 0.15. The relative fractions of the BCC, FCC, and HCP clusters are still very small. This result indicates the local orders around the Ni atoms in the glass sample are those similar to Mixed-ICO-Cube, Twined-Cube and ICO-like clusters.

In comparison with binary $Cu_{64.5}Zr_{35.5}$ which has much stronger GFA than $Ni_{64.5}Zr_{35.5}$, our MD simulation results show that the fraction of ICO-like cluster in $Ni_{64.5}Zr_{35.5}$ is much less than that in $Cu_{64.5}Zr_{35.5}$ [38]. It suggests that the ICO cluster indeed plays an important role in glass formation in intermetallic systems. The difference in the ICO fraction in the two systems can be attributed to the strong hybridization between Ni-Zr *d* electrons which is much weaker in Cu-Zr systems, as demonstrated by previous *ab* initio MD simulation studies [39, 40].

Chemical order is also a useful parameter to describe and predict the GFA in many alloys. To characterize the chemical order in our MD glass samples all the clusters with the alignment score < 0.15 to a given template are superposed by overlapping the center Ni atoms and keeping the orientation as that at the best alignment position. As shown in Figs. 6(a)-(c), one can observe that the collective alignment of these clusters gives a pattern of the template, where the red and blue spheres denote Ni and Zr atoms, respectively. In order to calculate the chemical occupation probability, we count the percentage of the Ni and Zr atoms at each site, after all the aligned clusters are overlap together. Fig. 6(d)-(f) are the calculated occupation probabilities of Ni or Zr on the sites of templates. For Mixed-ICO-Cube clusters, the atoms 2, 3, 4, 7, 8, 9, 10 can be counted as cube atoms. Our results show that the chemical composition at sites 2, 4, 8, 10 is Ni-rich (although it is a bit weak in site 8) while at sites 3, 7, 9 is Zr-rich, exhibiting the trend of B2 chemical order. At the other sites of the cluster, Ni and Zr concentrations are close to the stoichiometry of $Ni_{64.5}Zr_{35.5}$. This trend of B2 chemical order is even more pronounced in the Twined-Cube clusters as shown in Figs. 6(b) and (e). The weaker chemical order in the Mixed-ICO-Cube motif as compare to that in the Twined-Cube motif would be due to the mix of ICO structure in the Mixed-ICO-Cube motif. By contrast, the Ni and Zr concentrations at each site of the ICO-like clusters are close to the stoichiometry of $Ni_{64.5}Zr_{35.5}$, as one can see from Figs. 6(c) and (f). From this analysis, we can see that all the sites with pronounce chemical order are associated with the cubic corners of templates. These sites with excellent chemical order can be viewed as the "genes" of crystalline phase. The emergence of such crystalline genes even when the sample is quenched at very fast cooling rates could be responsible for the weak GFA in the $Ni_{64.5}Zr_{35.5}$ sample.

We also examined the MRO around the Mixed-ICO-Cube and Twined-Cube SRO clusters. By overlapping the center atoms of the same SRO clusters and keeping the orientation of the clusters at the positions from the SRO alignment with a given template, we can see the alignment distribution of the atoms in the second shell of these clusters. As shown in Fig. 7(a), we can see that the atoms in the second shell of the Mixed-ICO-Cube clusters resemble the B2 structure. In particular, one can observe that the second shell sites are mostly occupied by Ni atoms, indicating excellent chemical order of B2 crystal fragment. Similarly, the atoms of the second shell around the Twined-Cube cluster centers are also arranged in bcc-like lattice, as shown in Fig. 7(b). Like in the case of Mixed-ICO-Cube motif, we found the majority of the atoms around these bcc-lattice sites in the second shell of Twined-Cube clusters are Ni. These MRO results demonstrate a certain degree of crystal nucleation ability from Mixed-ICO-Cube and Twined-Cube cluster, suggesting the suppression of glass phase formation in $Ni_{64.5}Zr_{35.5}$ sample.

Recently, it has been demonstrated that performing MD simulated annealing below but close to the glass transition temperature (i.e., sub-$T_g$ annealing) can accelerate structural relaxation since the structural relaxation and glass formation is much more efficient in the vicinity of $T_g$ [38, 41, 42]. To investigate the effects of sub-$T_g$ annealing in Ni-Zr systems, the initial $Ni_{64.5}Zr_{35.5}$ sample at 2000K is continuously cooled down to 925K at $10^{10}$ K/s, then the as-quenched sample is annealed for 1400ns at 925K. Finally, the annealed sample was continuously cooled down to 300K at $10^{10}$ K/s. The MD steps used for such a cooling-annealing-cooling cycle is equivalent to a cooling rate of $1.08\times10^9$ K/s. Fig. 8(a) shows the potential energy at 300K for different cooling rates including result from the sub-$T_g$ annealing cycle. It can be seen that the potential energy for the samples with continuous cooling linearly depends on logarithm of cooling rates, which is similar to the case of other binary alloy systems such as Cu-Zr [41], Al-Sm [42] and Ni-Nb [43]. When considering the error bar of the fitting, the potential energy for $Ni_{64.5}Zr_{35.5}$ from the sub-$T_g$ annealing simulation is actually not far away from the extrapolation of the least-square fitting line based on the energies of the continuously cooled samples. This result suggests that in order to get the same potential energy, the simulation time used for sub-$T_g$ annealing is almost the same as that used in continuously cooling. This result is very different from the case of strong glass forming system like $Cu_{64.5}Zr_{35.5}$, where the sub-$T_g$ annealing approach can help to speed up the glass formation process [41]. Our results indicate that in weak glass forming system like the present case of $Ni_{64.5}Zr_{35.5}$, sub-$T_g$ annealing is not very effective owing to the relatively stronger crystalline nucleation tendency.

To study the influence of cooling rate on the SRO clusters, we show the fractions of Mixed-ICO-Cube, Twined-Cube and ICO-like clusters in the sample of 300K with different cooling rates in Fig. 8(b). When the cooling rate is lowered, the fraction of ICO-like clusters is nearly constant while that of Twined-Cube clusters increases considerably. The fraction of the Mixed-ICO-Cube clusters also increases slightly with lowering the cooling rate. The open symbols shown in Fig. 8(b) are the fractions of SRO clusters for the samples after sub-$T_g$ annealing process with an effective cooling rate of $1.08\times10^9$ K/s. Since the Twined-Cube cluster has the character of crystalline phase while ICO-like cluster is more related to glass phase, our results suggest that even in the very fast quench process the crystalline phase motifs are still the dominant SRO in $Ni_{64.5}Zr_{35.5}$ system. Therefore, $Ni_{64.5}Zr_{35.5}$ should not be a good glass former.

To further investigate the competition among different SRO motifs, we also examine the evolution of SRO clusters during the sub-$T_g$ annealing at 925K. During the annealing process, for every 200ns the sample is quenched to 10K to obtain the inherent structure for SRO analysis. Fig. 8(c) shows the evolution of fractions of Mixed-ICO-Cube, Twined-Cube and ICO-like clusters as the function of annealing time at 925K. One can observe that the fractions of Twined-Cube and ICO-like cluster fluctuate with annealing time. The fraction of the Mixed-ICO-Cube clusters is high, more than 17% at the beginning of the annealing and increases sharply during the first 400 ns to about 20%, and then reduces to about 18% from 500 to 1200 ns. At the same time, the fraction of the Twined-Cube clusters increase from about 10 to 11% during the last 400 ns of annealing. By tracking the evolution history of dominant motifs during the annealing simulation, we found that the change of fractions is a dynamic process and different types of motifs in the sample can interchange with each other in the process of the simulation. The fractions of Mixed-ICO-Cube and Twined-Cube clusters as the function of simulation time seen in Fig. 8 (c) is a statistical average result at each simulation time. As the annealing time gets longer and longer, more and more clusters are found to transfer into crystal-like Twined-Cube motif and reduce the GFA in the sample.

## IV. Conclusions

In summary, the atomic structure of $Ni_{64.5}Zr_{35.5}$ alloy has been studied by MD simulation combined with the cluster alignment method analysis. The calculated total structure factors agree well with the XRD and ND scattering experimental data. By pairwise cluster alignment and clique analysis, we found three kinds of Ni-centered motifs are the dominant SRO in $Ni_{64.5}Zr_{35.5}$ sample at 300K, i.e., Mixed-ICO-Cube, Twined-Cube and ICO-like cluster. The Mixed-ICO-Cube motif has half ICO and half cube structure. The Twined-Cube cluster is constituted of two interpenetrating slightly distorted cubes with common (111) plane. The ICO-like cluster is very close to ideal ICO with slight distortion. Meanwhile, the fractions of the other kinds of Ni-centered motifs and all Zr-centered motifs are much smaller than that of the above three dominant motifs.

In addition, we also analyze the chemical order of the three dominant SRO clusters by directly evaluating the occurrence probability of Ni (Zr) atoms around each sites of the SRO cluster. Our results show that the cubic portion of Mixed-ICO-Cube and the whole Twined-Cube clusters have excellent chemical order, implying the emergence of crystalline motifs in $Ni_{64.5}Zr_{35.5}$ alloy even in fast cooled samples. Furthermore, we studied the MRO of the Mixed-ICO-Cube and Twined-Cube clusters and found that around the bcc centers there is a high probability of finding another Ni atoms at the second shell from the central Ni atom. This suggests that the Mixed-ICO-Cube, Twined-Cube clusters have the character of crystalline metastable B2 phase in Ni-Zr system. We also found the fraction of Twined-Cube cluster increases with lowering cooling rate while that of ICO-like cluster remains constant, implying the glass formation would be suppressed by the small crystalline fragments of B2 phase. The evolution of SROs with the annealing time also shows competition between Twined-Cube and ICO-like clusters. Our results suggest that the weak GFA can be attributed to the existence of SROs with crystalline character in $Ni_{64.5}Zr_{35.5}$ alloy.

**Acknowledgements**

Work at Ames Laboratory was supported by the U.S. Department of Energy (DOE), Office of Science, Basic Energy Sciences, Materials Science and Engineering Division including a grant of computer time at the National Energy Research Supercomputing Center (NERSC) in Berkeley. Ames Laboratory is operated for the U.S. DOE by Iowa State University under contract # DE-AC02-07CH11358. This work was also supported by the National Natural Science Foundation of China (Grant Nos. 11304279 and 11104247). L. Tang would like to acknowledge the financial support from China Scholarship Council (No. 201608330083). T. Q. Wen and N. Wang would like to acknowledge the financial support from the National Natural Science Foundation of China (Grant Nos. 51671160 and 51271149).

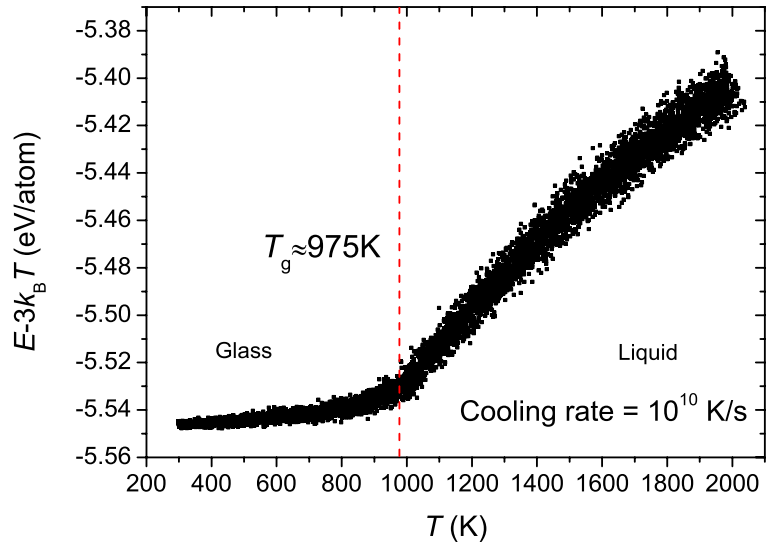

Fig. 1. The instantaneous energy $E$-$3k_BT$ as function of temperature with continuous cooling process, where the cooling rate is $10^{10}$ K/s. It is clear that there is a phase transition from liquid to glass and the obtained glass transition temperature $T_g \approx 975K$ agrees with the experimental data [19].

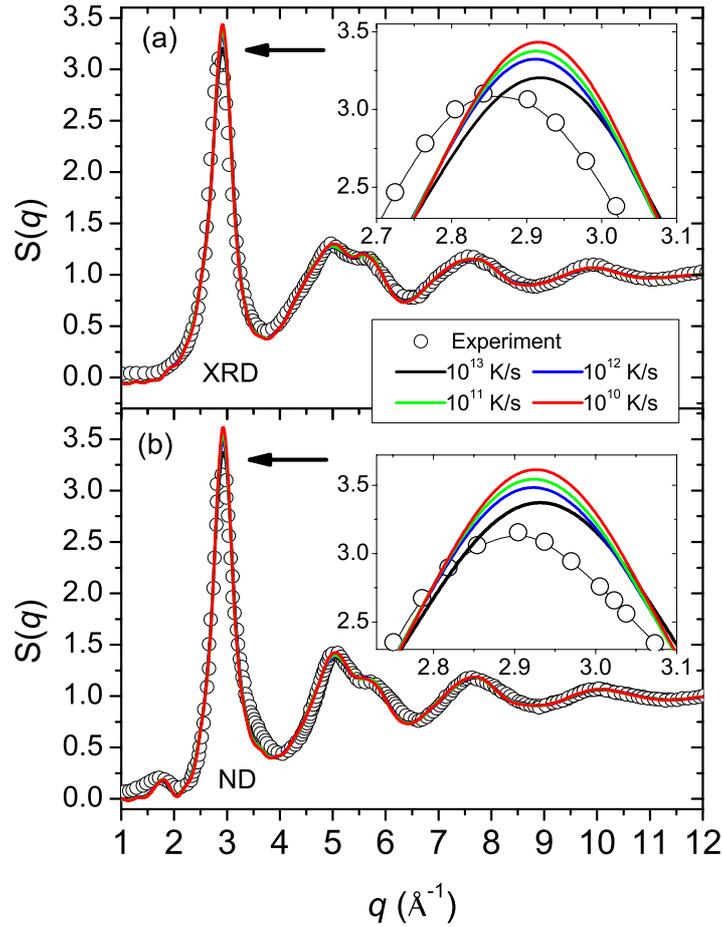

Fig. 2. The comparisons of total structure factors from the MD simulations to the (a) XRD and (b) ND experimental data [26], respectively. One can observe that the calculated total structure factors from MD simulations agree well with the XRD and ND experimental data. The maximal discrepancy between MD simulations and experiments locates at the main peak in low-$q$ region, as shown in the insets. In addition, the ND data from our MD simulation also produce well the characteristic pre-peak [25] around $q \approx 1.8$ Å$^{-1}$, which generally corresponds to medium-range correlations in real space.

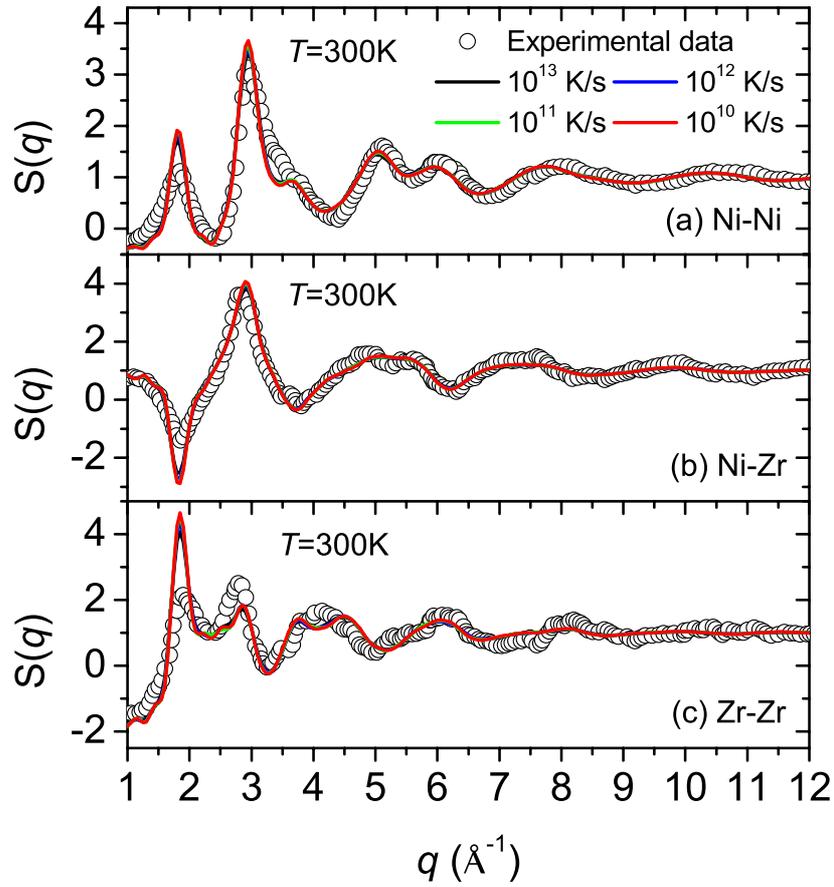

Fig. 3. Partial structure factors of (a) Ni-Ni, (b) Ni-Zr and (c) Zr-Zr for $Ni_{64.5}Zr_{35.5}$. In spite of some differences between calculated MD simulations and experimental data, the peaks of Ni-Ni and Ni-Zr around $q=3$ (Å$^{-1}$) corresponding to the dominant SRO clusters in $Ni_{64.5}Zr_{35.5}$ only have some minor deviations.

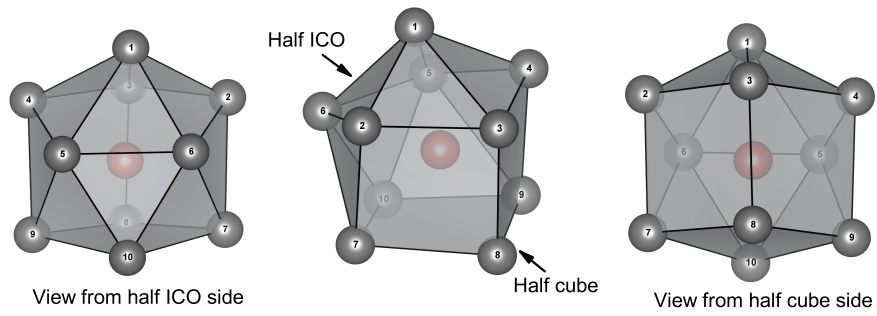

(a) Mixed-ICO-Cube cluster

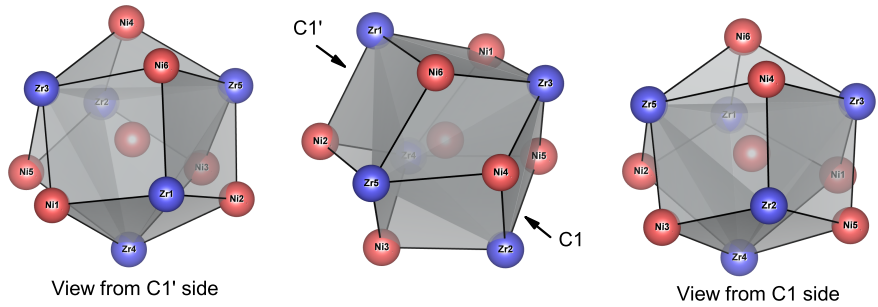

(b) Twined-Cube cluster

Fig. 4. Two kinds of dominant clusters extracted from the $Ni_{64.5}Zr_{35.5}$ glass sample at 300K by pairwise alignment method. (a) Mixed-ICO-Cube cluster can be viewed as the combination of half cube and half icosahedron surface. The sites (2, 3, 8, 7), (3, 4, 9, 8) and (7, 8, 9, 10) form three nearly orthogonal squares, respectively. The other portion of surfaces are 10 triangles constituting a slightly distorted half ICO. (b) Twined-Cube cluster can be viewed as two interpenetrating cubes with common (111) plane formed by sites (Zr3, Zr4, Zr5). The sites (Ni3, Ni4, Ni5, Zr2, Zr3, Zr4, Zr5) and (Ni1, Ni2, Ni6, Zr1, Zr3, Zr4, Zr5) form the two half cubic surfaces C1 and C1', respectively. Here the usage of Ni or Zr as the site's name is to imply that the corresponding sites could be mostly occupied by Ni or Zr as revealed in the following investigation.

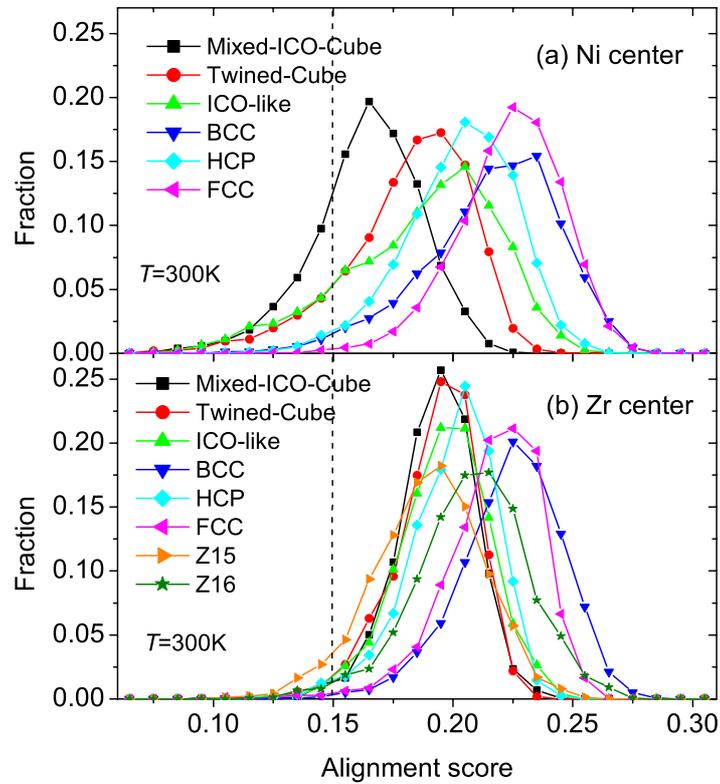

Fig. 5. The distributions of alignment scores against various motifs for (a) Ni- and (b) Zr-centered clusters in glass sample at 300K. It can be seen that the score distributions of Ni-centered Mixed-ICO-Cube, Twined-Cube, and ICO-like motifs have relatively larger portion in the region of small score, indicating that these three motifs are dominant SRO clusters in $Ni_{64.5}Zr_{35.5}$. Choosing a cutoff score value of 0.15, we can obtain the fractions of Mixed-ICO-Cube, Twined-Cube, and ICO-like clusters are 18.6%, 9.3% and 11.8% of the total Ni atoms, respectively.

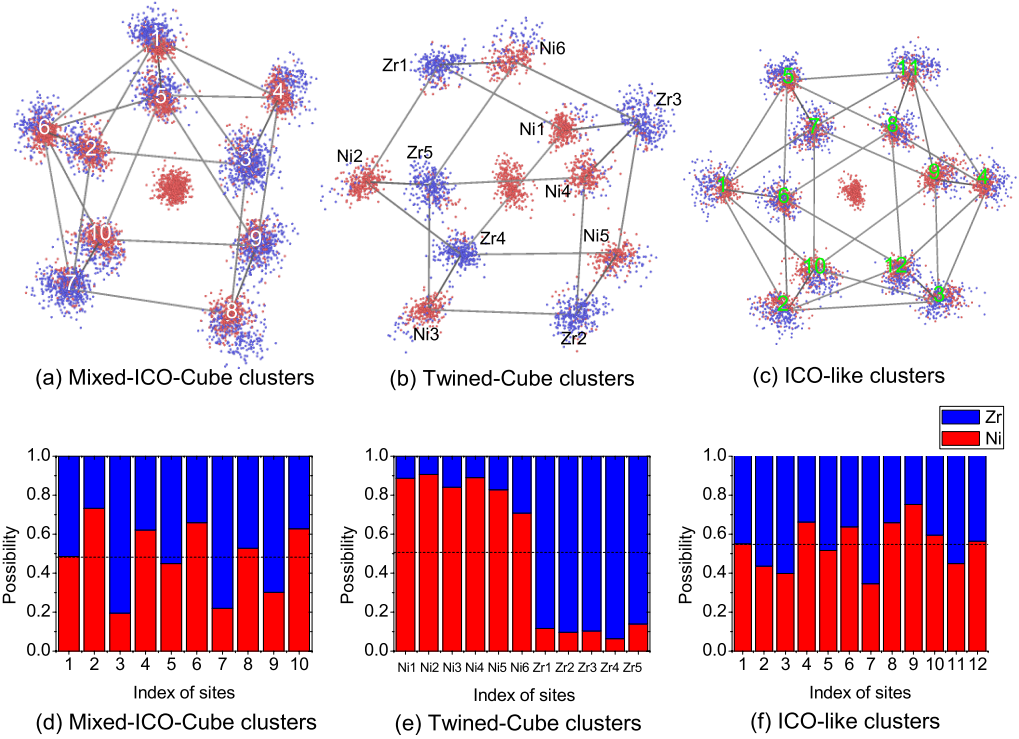

Fig. 6. The atomic configurations of collective alignment for (a) Mixed-ICO-Cube, (b) Twined-Cube and (c) ICO-like clusters, where the red and blue sphere denote Ni and Zr atoms, respectively. The possibility of each site being Ni (red) or Zr (blue) is also shown in (d)-(f), where the dashed lines denote the averaged possibilities. In the cubic part of Mixed-ICO-Cube clusters, the chemical composition at sites 2, 4, 8, 10 is Ni-rich (although it is a bit weak in site 8) while at sites 3, 7, 9 is Zr-rich, exhibiting the trend of B2 chemical order. This trend of B2 chemical order is even more pronounced in the Twined-Cube clusters. Our results demonstrate that the crystalline-like clusters have much more pronounced chemical order than that of ICO, implying that the crystalline phase could still emerge in $Ni_{64.5}Zr_{35.5}$ sample even by fast cooling process.

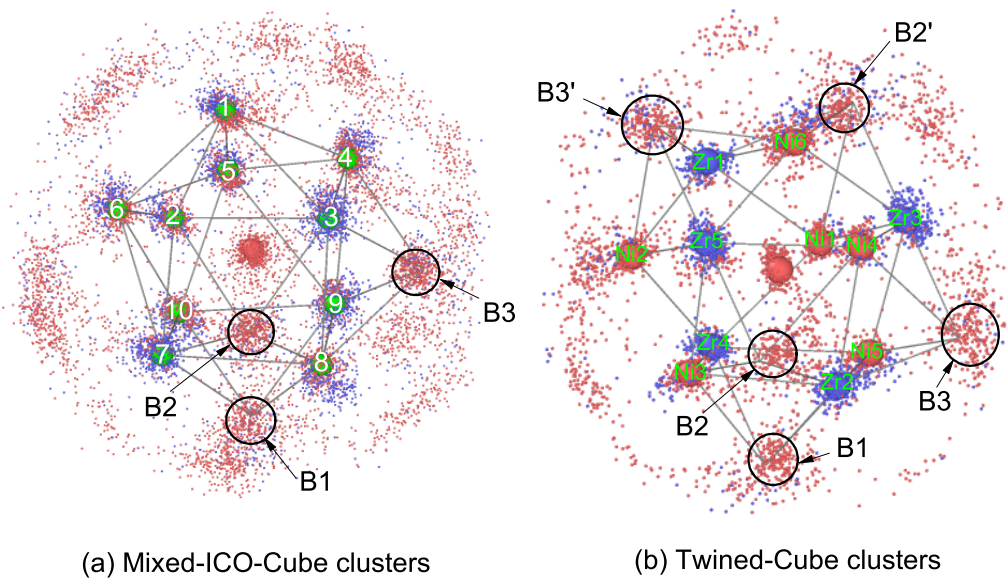

Fig. 7. The atomic configurations of collective alignment of (a) Mixed-ICO-Cube clusters and (b) Twined-Cube clusters with the second shell atoms. For Mixed-ICO-Cube clusters, one can see that there are accumulations of the second shell atoms (mostly Ni, red sphere) around the B1, B2 and B3 positions which can be viewed as three nearly bcc centers. It suggests that the half cubic part of Mixed-ICO-Cube cluster has the trait of metastable B2 phase in Ni-Zr system. Similar to the case of Mixed-ICO-Cube clusters, for Twined-Cube clusters the possibility of finding a Ni atom around the bcc centers (B1, B2, B3, B2', B3') of two interpenetrating cubes (C1, C1') is relatively larger, also implying the existence of metastable B2 phase in $Ni_{64.5}Zr_{35.5}$.

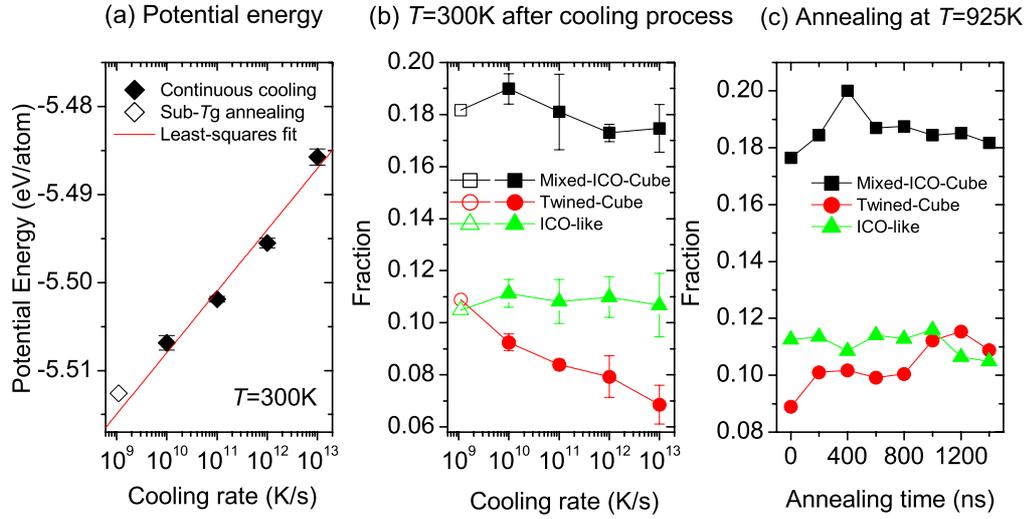

Fig. 8. (a) The potential energy of $Ni_{64.5}Zr_{35.5}$ at 300K for different cooling rates. (b) The fractions of Mixed-ICO-Cube, Twined-Cube and ICO-like clusters in $Ni_{64.5}Zr_{35.5}$ at 300K for different cooling rates. With lowering cooling rate, the fraction of Twined-Cube cluster which represents crystalline phase increases while the fraction of ICO-like cluster is nearly constant, suggesting the suppression of glass phase formation by the competitors (crystalline phase) in $Ni_{64.5}Zr_{35.5}$ sample. (c) The evolution of fractions of Mixed-ICO-Cube, Twined-Cube and ICO-like clusters during the annealing process at temperature 925K. One can observed that there is clear phenomenon of competition between the representation of crystalline phase (Twined-Cube cluster) and the glass phase (ICO-like cluster).

Table. I The fractions of SROs in the glassy sample at 300K with continuous cooling rate $10^{10}$ K/s, where the cutoff of alignment score is 0.15.

|  | BCC | FCC | HCP | Mixed-ICO-Cube | ICO-like | Twined-Cube | Z15 | Z16 |
| --- | --- | --- | --- | --- | --- | --- | --- | --- |
| Ni-centered | 0.9% | 0.2% | 1.2% | 18.6% | 11.8% | 9.3% | N/A | N/A |
| Zr-centered | 0.2% | 0.8% | 1.5% | 1.0% | 1.7% | 1.0% | 4.8% | 1.5% |